# SAGE: Synthetic Aging for a Grid Environment

Paolo Esquivel[1], Mark A. Harris[2], and Stephen J. Harris[3]

[1]Independent Consultant, [2]Independent Consultant, [3]Lawrence Berkeley National Lab, Berkeley, Ca 94720

**ABSTRACT**

Grid-scale battery degradation unfolds over multi-year timescales under coupled electrochemical, thermal, and operational feedbacks that are difficult to capture using laboratory cycling data or proprietary field datasets. This limits the development and validation of degradation-aware algorithms, fleet management strategies, and digital twin methodologies that require long-horizon, physically consistent ground truth.

Here we present SAGE (Synthetic Aging for a Grid Environment)**,** an open-source, physics-informed simulation framework that generates hour-resolved, multi-decade operating histories and degradation trajectories for heterogeneous battery energy storage system (BESS) fleets under realistic dispatch. The framework couples stochastic environmental drivers, market-based dispatch logic, electro-thermal behavior, calendar and cycle aging kinetics, and asset-level heterogeneity within a transparent, externally parameterized architecture designed for reproducibility and systematic sensitivity analysis.

We validate the physical consistency of the framework through hierarchical tests, including verification of Arrhenius temperature acceleration, correct propagation of thermal stratification effects, emergent wear-out failure statistics, and internally consistent coupling between efficiency fade, heat generation, and degradation accumulation. Fleet simulations demonstrate how intrinsic heterogeneity in thermal environment and manufacturing quality naturally produces dispersion in state-of-health trajectories and asset lifetimes, without imposing statistical assumptions on failure distributions.

SAGE is intended as a general-purpose synthetic data generator and benchmarking platform for degradation-aware optimization, state estimation, machine learning, and fleet-scale risk analysis, without attempting asset-level lifetime forecasting. By providing open, configurable, and physically grounded long-horizon datasets, the framework enables controlled exploration of battery aging behavior in BESS and supports reproducible research in grid-scale energy storage modeling.

**HIGHLIGHTS**

- Open-source physics-informed framework for multi-decade BESS fleet simulation
- Couples aging kinetics, thermal feedback, stochastic environments, and dispatch
- Hierarchical validation confirms Arrhenius-consistent calendar and cycle aging
- Synthetic datasets with controllable heterogeneity and sensor noise for benchmarking
- Naive lifetime models can match EOL yet lack fleet-level variability

# 1. INTRODUCTION

Understanding long-term degradation behavior of grid-scale battery energy storage systems (BESS) is essential for the development of reliable control strategies, health estimation algorithms, and fleet management methodologies. However, the data required to support these efforts remains limited. Laboratory aging experiments on (almost always) single cells provide high-fidelity electrochemical insight, but they are generally constrained by simplified duty cycles, and they are blind to emergent degradation mechanisms that involve multiple batteries in a heterogeneous environment. Conversely, field data from deployed systems capture realistic operational variability but are often proprietary, incomplete, and lack the ground truth required for systematic algorithm validation. These limitations create a persistent gap between physically rigorous modeling and data-driven development for long-horizon grid applications. The growing national emphasis on AI-enabled energy infrastructure research, reflected in recent U.S. Department of Energy science and technology challenge announcements[1], further underscores the need for reproducible, physics-grounded simulation platforms that support algorithm development and benchmarking in grid-scale storage.

Synthetic data generation has emerged as a promising approach for bridging this gap. By producing large volumes of labeled operating data under controlled assumptions, synthetic frameworks enable repeatable benchmarking, sensitivity analysis, and algorithm stress testing that are difficult or impossible to perform using laboratory or field datasets alone. Existing synthetic data generators across the power and energy domain span a wide range, from grid-centric load and network simulators to data-driven trajectory generators based on historical replay or machine learning. While valuable within their respective scopes, many existing tools abstract away internal battery physics, thermal feedback, and degradation dynamics, limiting their suitability for applications that require physically consistent aging behavior over multi-year horizons.

Recent digital twin efforts have advanced the integration of electrochemical and thermal models for individual battery systems, enabling improved state estimation and asset-level optimization. However, extending such high-fidelity models to large fleets and long time horizons remains computationally challenging and often requires proprietary calibration data[2, 3]. Moreover, many modeling approaches rely on representative-asset assumptions that neglect intrinsic heterogeneity arising from manufacturing variability, spatial thermal gradients, and stochastic operating histories[4, 5]. In real deployments, these sources of variability produce distributions of degradation trajectories and asset lifetimes rather than a single deterministic path, motivating the need for simulation tools that explicitly represent population-level behavior[6].

To address these challenges, we introduce SAGE (Synthetic Aging for a Grid Environment), an open-source, physics-informed simulation framework designed to generate long-horizon synthetic operating histories and degradation trajectories for heterogeneous BESS fleets under realistic dispatch. The framework couples stochastic environmental drivers, market-based scheduling, electro-thermal behavior, calendar and cycle aging kinetics, and asset-level heterogeneity within a modular and externally parameterized architecture. All model parameters are supplied through validated configuration files to promote transparency, reproducibility, and systematic exploration of uncertainty.

Because degradation reflects asset-specific operating histories, environmental exposure, and dispatch decisions, lifecycle behavior cannot be reliably inferred from aggregate system models alone. Accordingly, the primary objective of SAGE is not to predict the behavior of any specific deployed asset, but to provide a controlled simulation testbed for developing and benchmarking degradation-aware algorithms, machine learning models, and fleet-scale analysis methods prior to deployment[7, 8]. The framework produces internally consistent latent state trajectories together with sensor-corrupted observations, enabling controlled evaluation of estimation, control, and diagnostic workflows under realistic noise and operational variability[9, 10].

This paper describes the architecture of the SAGE framework, documents the governing physical models and abstractions, and validates the implementation against theoretical expectations and emergent statistical behavior. We demonstrate that the coupled electro-thermal and degradation models reproduce correct Arrhenius temperature acceleration, capture the effects of thermal stratification, and generate realistic wear-out–dominated lifespan distributions in heterogeneous fleets. The modular and fully parameterized design of SAGE allows straightforward extension to incorporate additional physics, operating regimes, and application-specific degradation mechanisms. The framework is released as open-source software to encourage reproducibility, extension, and community-driven development in grid-scale energy storage research.

## 2. LITERATURE REVIEW

The scarcity of open-source, long-horizon battery degradation data under realistic grid operation has motivated the development of a range of synthetic data generators (SDGs) across the energy domain. Existing SDGs vary widely in scope, from tools that replicate electrical boundary conditions to models that augment observed operational data. However, most approaches are optimized either for grid-level realism or for electrochemical fidelity, rarely both simultaneously. As a result, they are generally ill-suited for generating degradation-labeled datasets that reflect the coupled effects of stochastic markets, environmental variability, and battery physics over multi-year horizons, the regime most relevant for grid-scale battery energy storage systems (BESS)

### 2.1 Grid-Centric and Load Generators

A significant body of work focuses on synthesizing realistic electrical loads and network topologies to validate power flow algorithms. Krishnan et al.[11] developed a framework for validating synthetic distribution system data, using statistical metrics to ensure that synthetic feeders match the topological complexity of real utilities. Lahariya et al.[12] introduced an SDG for electric vehicle charging sessions, utilizing an exponential distribution to model arrival times, and Gaussian mixture models to simulate connection durations and required energy based on real-world patterns.

While valuable for testing power flow and market-clearing mechanisms, these tools treat energy storage assets as static nodes with fixed efficiency and no internal state evolution[11, 13]. Such grid-centric SDGs cannot produce state-of-health labels or degradation trajectories, as the internal electro-thermal physics of the battery are abstracted away. This limitation precludes their use for developing and benchmarking degradation-aware dispatch, asset-preservation strategies, or maintenance-oriented analysis methods that require physically consistent aging behavior[14, 15].

## 2.2 Historical Replay and Clustering

To capture temporal dynamics of grid storage, researchers may turn to historical market data. Moy et al.[16] demonstrated a methodology using K-means clustering and Principal Component Analysis to synthesize representative duty cycles from real-world peak shaving dispatch profiles, ensuring that frequency content and power volatility are empirically grounded. However, replay-based models are inherently open-loop: they assume that the battery's response to a given signal remains invariant over time.

In practice, degradation alters internal resistance, efficiency, and thermal behavior, which in turn modifies the battery's ability to follow dispatch signals and accelerates further aging[3]. By decoupling signal generation from the battery's evolving physical state, replay models omit these feedback loops, limiting their ability to reproduce late-life behavior and compounding degradation effects observed in operating systems.

## 2.3 Data-Driven Augmentation

Data-driven approaches using Generative Adversarial Networks (GANs) and Variational Autoencoders (VAEs) have emerged for load profiling and battery trajectory synthesis[17, 18]. These approaches can efficiently interpolate within observed data distributions and have shown promise for trajectory completion and short-horizon forecasting. However, because they are non-mechanistic, they encode statistical correlations rather than physical laws. As a result, purely data-driven generators struggle to produce physically meaningful counterfactual scenarios, such as alternative dispatch strategies, extreme weather events, or market rule changes that fall outside the training distribution[17, 19] For applications involving degradation-aware optimization or long-term asset planning, the inability to extrapolate beyond historical operating regimes is a fundamental limitation[20].

## 2.4 Laboratory Datasets

Laboratory cycling remains the gold standard for electrochemical rigor, but controlled laboratory protocols do not capture the stochastic thermal gradients and dispatch variability characteristic of grid operations[21, 22]. Such studies cannot exhibit the emergent heterogeneity that arises when manufacturing variability[23, 24], thermal stratification[3, 25], and variable signal structure interact over years of operation

Field studies have documented these real-world complexities, including thermal gradients as high as 9°C within single installations[26] and capacity spread increasing over time[21], but observational data cannot provide the controlled counterfactual comparisons needed to isolate parameter effects or validate dispatch algorithms.

## 2.5 Optimization Frameworks and Digital Twins

State-of-the-art optimization frameworks have successfully integrated semi-empirical aging penalties to internalize cycling costs, enabling degradation-aware dispatch[8, 27, 28]. Recent physics-based digital twin studies by Movahedi et al.[29] have introduced metrics such as "throughput gained versus days lost," demonstrating that for batteries dominated by calendar aging, the economic utility of grid services often outweighs marginal degradation cost. However, these models are typically applied to individual assets or representative systems, limiting their ability to capture population-level effects such as tail risk[6], early failures, and lifespan dispersion across fleets. Extending such high-fidelity models to large ensembles remains computationally challenging, motivating the need for synthetic frameworks that preserve physical structure while enabling scalable fleet-level analysis.

A key distinction separates calibration-driven digital twins from scenario-generating synthetic engines. Digital twins are calibrated to specific fielded assets using operational data and track real-time state evolution for a particular system; their value lies in asset-specific prediction and optimization. SAGE serves a complementary purpose: rather than modeling any specific asset, it generates unlimited physically allowed counterfactual scenarios, varying market conditions, thermal environments, fleet heterogeneity, and degradation parameters, to support algorithm development, sensitivity analysis, and fleet-level risk quantification before deployment. This distinction positions SAGE as a pre-deployment research tool rather than an operational monitoring system.

## 2.6 The Representative Agent Problem

A systemic limitation appears in some existing SDGs: reliance on representative agent modeling[30]. Tools typically generate data for a single idealized asset, a golden sample or identical clones, rather than a statistical population[31]. This approach fails to capture the intrinsic heterogeneity of grid-scale fleets, where manufacturing tolerances[24], staggered commissioning dates[21], and uneven thermal management[4] create wide distributions of state-of-health trajectories. Spatial temperature gradients within utility-scale containers[26] can produce significant capacity spreads between battery packs over time, yet homogeneous models predict zero dispersion.

Validation on homogeneous data risks underestimating tail risk, as it ignores the weakest link[3, 4] dynamics that often dictate reliability of large-scale storage systems. For warranty analysis[6, 20, 32], maintenance optimization, and fleet management, understanding the full distribution of outcomes, not just the mean trajectory, is essential.

These limitations motivate a synthetic data generation framework that treats degradation, heterogeneity, and fleet dispersion as emergent properties of coupled physics and stochastic operation rather than imposed assumptions. SAGE is designed to address this gap.

## 2.7 Synthetic electrochemical data

Synthetic electrochemical data has become a widely used tool for developing and stress-testing[33] battery state and health estimation methods, particularly when long-horizon field datasets are sparse, proprietary, or do not span the operating regimes of interest. Early studies demonstrated the value of synthetic signals for validating state-of-health estimators and diagnostic methods[34], and subsequent work released large public synthetic datasets to support algorithm training and benchmarking, often under constant-current or tightly controlled laboratory protocols[35]. More recent studies have emphasized physics-based synthetic generation, pack-level effects, and experimental validation of synthetic-data utility for inference and decision-making[23, 36, 37]. SAGE builds on this direction by generating long-horizon, fleet-resolved synthetic operational histories that couple prices, dispatch, electro-thermal response, and aging, enabling controlled sensitivity studies and benchmarking under grid-relevant constraints rather than fixed-current laboratory regimes.

## 3. METHODS

### 3.0 Motivation and Design Objectives

SAGE, introduced in this work, is an open-source, physics-informed simulation framework that generates hour-resolved[38], multi-decade synthetic operating histories and degradation trajectories for heterogeneous BESS fleets under realistic dispatch. The framework supports four complementary uses: synthetic data generation for machine-learning workflows, ground-truth benchmarking of state-estimation algorithms, characterization of fleet-level heterogeneity and population-level variability, and exploratory evaluation of dispatch and control strategies using physically grounded degradation outcomes.

### 3.1 Design Philosophy

Three design principles guide the architecture:

*3.1.1. Separation of scheduling and physics*. Market decisions determine when discharge occurs, while the physics engine determines how much power can actually be delivered. This separation ensures that dispatch commands reflect realistic operator decisions, while physical constraints, thermal limits, efficiency losses, and capacity fade, emerge from underlying electrochemical behavior rather than ad hoc control rules.

*3.1.2. Explicit parameterization*. All model parameters are supplied through external configuration files with mandatory validation. Missing parameters cause immediate failure rather than silent substitution of defaults. This fail-fast philosophy ensures reproducibility, facilitates systematic sensitivity analysis, and prevents hidden assumptions from contaminating results.

*3.1.3. Ground truth with measurement realism*. SAGE exports both true physical states (as computed by the physics engine) and sensor-corrupted measurements with configurable noise levels. This dual-output design enables validation of state-estimation and control algorithms against known ground truth while training and testing on realistically noisy observations.

SAGE generates plausible counterfactual operating histories rather than forecasts of specific assets. The framework is designed for scenario exploration: the baseline parameters presented here represent one working configuration that produces physically plausible outputs, not universal defaults. Researchers studying various chemistries, markets, climates, or dispatch strategies should substitute parameter values appropriate to their application.

**3.2 Scope and Abstractions**

To preserve transparency and generality, several elements are intentionally abstracted in the current implementation:

*3.2.1 Charging physics and grid import costs.* Discharge behavior in grid-scale BESS is governed by well-defined market products, typically fixed-duration blocks at approximately constant power, yielding operating profiles that are broadly comparable across deployments[39]. Charging protocols, by contrast, depend on asset-specific contractual, infrastructural, and co-location relationships, including collocated renewables, congestion constraints, curtailment capture, and grid import limits, that vary fundamentally across systems[40, 41].

SAGE therefore models discharge physics explicitly while treating charging as an exogenous boundary condition. All degradation coefficients are interpreted as effective full-cycle aggregates that incorporate stress accumulated during both charge and discharge, rather than as discharge-only physics[22, 28]. This abstraction preserves physical consistency at the cycle level while avoiding the introduction of asset-specific charging assumptions—such as isolated microgrid load-following or rigid SoC-return rules—that would reduce generality. This assumption is discussed further in section 4.8.3.

*3.2.2 Spatial electrochemistry within cells.* SAGE models cell-level behavior without resolving internal electrode-scale gradients.

*3.2.3 Sub-hourly transient dynamics.* The hourly timestep is appropriate for economic analysis[42] but does not resolve fast thermal transients. Transient thermal behavior depends on system-specific characteristics, such as thermal mass, airflow geometry, HVAC sizing, and pack placement, that vary substantially across deployments. Rather than impose arbitrary assumptions about these parameters[3], SAGE employs a quasi-steady-state thermal approximation, with limitations discussed in Section 3.7.4.

*3.2.4 Ancillary service market participation.* The current implementation focuses on energy arbitrage; ancillary services would require additional market models.

*3.2.5 Cell-to-cell interactions within packs.* Fleet heterogeneity in SAGE arises from manufacturing variability and thermal position, not from explicit modeling of series/parallel cell interactions[4, 5].

*3.2.6 Chemistry scope.* SAGE is designed as a chemistry-agnostic framework at the structural level: the coupling of stochastic environments, dispatch logic, electro-thermal feedback, and degradation accumulation does not assume a specific cell chemistry. The governing architecture therefore applies broadly across lithium-ion and related battery technologies. The baseline parameterization used for validation, however, is broadly consistent with lithium iron phosphate (LFP) behavior under moderate C-rates and temperatures. Application to other chemistries may require re-parameterization of degradation coefficients, stress functions, activation energies, and thermal limits. The baseline values presented here should be interpreted as illustrative rather than universal defaults.

## 3.3 Simulation Architecture

SAGE comprises four coupled modules executed sequentially for each simulated asset, with information passed forward at each timestep to capture feedback between market conditions, operational decisions, and degradation physics:

1. *Stochastic Environment Generator* produces synthetic hourly time series of ambient temperature and electricity prices over the full simulation horizon, incorporating realistic seasonal structure, diurnal patterns, and heavy-tailed scarcity events.
2. *Block Dispatch Scheduler* determines daily discharge timing using forecast prices, enforcing fixed-duration, block-constant operation without hour-by-hour re-optimization.
3. *Physics Engine* computes state-of-charge evolution, cell temperature, discharge efficiency, degradation accumulation, and safety-driven power derating at each timestep.
4. *Fleet Monte Carlo Layer* introduces intrinsic heterogeneity across assets through manufacturing quality variation and rack-position assignment, enabling fleet-scale analysis of degradation dispersion, asset-to-asset variability, and population-level lifespan statistics.

Prior to system-level validation, core physics functions are verified through 45 unit-tests comparing computed values against analytical expectations. Tests confirm correct Arrhenius factor behavior at reference temperature, proper degradation progress mapping, SOC window contraction with aging, and cycle aging proportionality to throughput. All tests pass within numerical tolerance.

Each module is independently auditable and governed by externally supplied parameters. The simulation advances at hourly resolution with explicit calendar structure. All temporal calculations assume a fixed 365-day year (8,760 hours). Leap days are omitted, causing calendar timestamps to drift approximately one day every four years relative to simulation time—a cumulative shift of six days over 25 years. This drift is negligible for physical quantities but necessitates simulation-time-based indexing when aggregating to monthly resolution. The following sections detail each module in sequence, beginning with the stochastic environment generator.

**3.4 Stochastic Environment Generation**

To evaluate degradation under realistic grid conditions, SAGE generates synthetic hourly time series for ambient temperature and electricity prices that preserve the statistical structure of real-world markets. Ambient temperature is modeled as a superposition of deterministic seasonal and diurnal harmonics with stochastic weather variability. This temperature series serves as a driver for both cell thermal dynamics and electricity price deviations. While temporally correlated stochastic processes (e.g., Ornstein–Uhlenbeck noise) may better represent short-term weather and price persistence, degradation outcomes in SAGE are governed by cumulative thermal and cycling stress integrated over months to years, a timescale over which short-range autocorrelation is not expected to materially influence lifetime statistics. Uncorrelated noise was therefore retained to preserve simplicity and to avoid introducing correlation-time parameters that cannot be independently identified from long-horizon degradation observations.

Electricity prices are generated using a coupled structural model. A deterministic "backbone" captures seasonal and diurnal load patterns, while weather-driven demand is modeled via heating and cooling degree-day sensitivities. To capture the heavy-tailed volatility characteristic of energy markets, we implement a scarcity mechanism that applies a conditional Pareto-distributed multiplier to base prices during extreme events. Real-world battery revenues are frequently dominated by rare scarcity events rather than average spreads. For example, recent market benchmarks show that a two-day cold snap in MISO created revenue opportunities of approximately $60/kW-month for a 4-hour battery, illustrating the economic importance of heavy-tailed price distributions[43]. Finally, to simulate dispatch under forecast uncertainty, the scheduler operates on a parallel forecast price signal generated from the same stochastic drivers as the realized prices but with independent noise. Full governing equations and calibration details for the temperature, price, and forecast models are provided in Supplementary Information S2.

All prices generated by SAGE are unit-consistent but inflation-agnostic. Users may interpret them as nominal or real depending on downstream modeling choices.

## 3.5 Dispatch Abstraction

SAGE employs a fixed-duration, constant-power discharge abstraction representative of common grid-scale energy arbitrage and resource adequacy products. Each simulated asset executes at most one discharge block per calendar day, with block timing selected based on forecast prices and held fixed for the full duration. No hour-by-hour re-optimization is performed, and dispatch decisions do not account for degradation state or physical constraints.

The dispatch module determines *when* discharge is requested, while the physics engine determines *how much* power can actually be delivered, subject to state-of-charge limits, efficiency losses, thermal constraints, and degradation-induced derating. This separation ensures that operational decisions remain market-driven, while physical feasibility and aging behavior emerge from the coupled electro-thermal and degradation models rather than from embedded control logic.

## 3.6 State of Charge Dynamics

Time development of SOC is given by Equation (1). Here, $P_{\text{grid}}(t)$ is the power delivered to the grid during discharge, $\eta_{\text{dis}}(t)$ is discharge efficiency, and $E_{\text{cap}}(t)$ is the current usable energy capacity (kWh) used to normalize SOC.:

$$\text{SOC}(t + \Delta t) = \text{SOC}(t) - \frac{P_{\text{grid}}(t)\,\Delta t}{\eta_{\text{dis}}(t)\,E_{\text{cap}}(t)} \tag{1}$$

where $P_{\text{batt}}(t) = \frac{P_{\text{grid}}(t)}{\eta_{\text{dis}}(t)}$

Here and throughout, the timestep is one hour, so power and energy are expressed on a consistent hourly basis. During non-dispatch hours $P_{grid}(t) = 0$, and therefore $P_{batt}(t) = 0$, so SOC remains unchanged. Each day begins at $\text{SOC}_{\max}(t)$ under the discharge-only abstraction described in Section 3.2.1.

As state of health declines, the allowable SOC operating window contracts toward end-of-life limits:

$$\text{SOC}_{\max}(t) = \text{SOC}_{\max,\text{BOL}} - (\text{SOC}_{\max,\text{BOL}} - \text{SOC}_{\max,\text{EOL}}) \cdot \frac{1-\text{SOH}(t)}{1-\text{SOH}_{\text{EOL}}} \qquad (2)$$

$$\text{SOC}_{\min}(t) = \text{SOC}_{\min,\text{BOL}} + (\text{SOC}_{\min,\text{EOL}} - \text{SOC}_{\min,\text{BOL}}) \cdot \frac{1-\text{SOH}(t)}{1-\text{SOH}_{\text{EOL}}} \qquad (3)$$

These equations shrink the usable SOC range as the battery ages. In SAGE, this is a compact representation of voltage/power feasibility under constant-power operation: increasing internal resistance raises the current required to meet a fixed power request and increases polarization losses ($IR$ drop), so voltage limits are encountered earlier within the nominal SOC range. Under baseline parameters, the allowable SOC window narrows from 0.05–0.95 at beginning of life to 0.20–0.80 at end of life[14, 44].

### 3.7 Thermal Model

The ambient temperature $T_{amb}$ represents the air temperature inside the battery container, which is regulated by HVAC. Container temperature is modeled as:

$$T_{amb}(t) = T_{setpoint} + \alpha \cdot (T_{outdoor}(t) - \overline{T_{outdoor}}) + \varepsilon(t) \qquad (4)$$

where $T_{setpoint}$ is the HVAC target (22°C baseline), α is the attenuation factor representing the fraction of outdoor temperature deviation that penetrates the container (0.0833 baseline), and $\varepsilon(t)$ is small-amplitude noise capturing HVAC control imprecision. With these parameters, a 46°C outdoor swing (−2°C to 44°C) produces only a 5–6°C container swing, consistent with well-designed thermal management systems. Setting α = 0 models perfect HVAC; larger values model degraded or undersized cooling.

Cell temperature determines degradation rates through Arrhenius kinetics. SAGE employs a quasi-steady-state thermal model:

$$T_{cell}(t) = T_{amb}(t) + \Delta T_{static} + \Delta T_{oper}(t) \tag{5}$$

### 3.7.1 Static Thermal Offset

Battery packs at different vertical positions within a rack experience different temperatures due to heat rising within the enclosure[26]. Packs near the top run hotter than those near the bottom. SAGE models this effect as:

$$\Delta T_{static} = g \cdot \Delta T_{gradient} \tag{6}$$

where $g$ represents vertical position within the rack (0 = bottom, 1 = top) and $\Delta T_{gradient}$ is the total temperature difference from bottom to top. For example, with a configured gradient of 5°C, a pack at the top of the rack ($g = 1$) operates 5°C warmer than a pack at the bottom ($g = 0$), while a pack at mid-height ($g = 0.5$) operates 2.5°C warmer.

### 3.7.2 Operating-Dependent Heating

When a battery discharges, not all stored energy reaches the grid; a fraction is dissipated as heat due to internal resistance. The lower the discharge efficiency, the more heat is generated. SAGE converts this heat into a temperature rise[26]:

$$\Delta T_{oper}(t) = K_T \cdot P_{heat}(t) \tag{7}$$

$$P_{heat}(t) = P_{grid}(t) \cdot \left(\frac{1}{\eta_{dis}(t)} - 1\right) \tag{8}$$

Equation (7) relates heat generation rate to temperature rise through $K_T$, an effective steady-state thermal gain (°C/kW) that aggregates enclosure heat rejection under quasi-steady operating conditions. Equation (8) calculates the heat generation rate from the difference between power drawn from the battery and power delivered to the grid. (This is where the feedback loop comes from: as efficiency drops, heat generation increases, which raises cell temperature, which accelerates Arrhenius-driven degradation, which further reduces efficiency.)

The thermal gain coefficient is not specified directly as $K_T$ in the configuration file. Instead, the user supplies a calibration parameter (calibrated_temp_rise_c4) that defines the steady-state temperature rise during C/4 discharge at beginning-of-life efficiency, 2º C in the baseline. The effective $K_T$ used in the quasi-steady thermal model is derived from this calibration so that the specified temperature rise is reproduced under nominal operating conditions. This calibration parameter is externally configurable, ensuring that enclosure thermal behavior can be adjusted without modifying the governing equations.

### 3.7.3 Thermal Safety Constraint

A hard safety limit is enforced: $T_{\text{cell}}(t) \leq T_{\text{cell,max}}$. If this constraint would be violated, power is derated at block initiation to maintain cell temperature below the limit[45].

### 3.7.4. Thermal Time Constants and Transients.
The thermal model is quasi-steady-state at the hourly timestep: operating heat generation produces an instantaneous temperature rise $\Delta T_{\text{oper}}(t)$ without explicit thermal inertia. This approximation is appropriate for the baseline operating regime (one 4-hour block per day at approximately C/4), where discharge duration substantially exceeds typical enclosure thermal time constants and temperatures approach steady-state within each hour. Unresolved transients, cooler temperatures at discharge onset, warmer temperatures during post-discharge cooling, likely produce a modest net underestimate of cumulative thermal stress due to the complexity of Arrhenius kinetics, but this effect is expected to be small relative to persistent factors such as ambient temperature and rack stratification. For higher C-rates, clustered dispatch events, or systems with limited thermal management, a thermal mass model may be required to capture transient heat retention.

### 3.7.5 Charge-Phase Thermal Effects.
The discharge-only abstraction omits heat generation during charging. In real systems, charge-phase heating elevates cell temperature at discharge onset, particularly for aggressive duty cycles with multiple daily cycles or short rest periods between charge and discharge. Under the baseline dispatch

(one 4-hour block per day with overnight charging), the ~20-hour interval between discharge events allows substantial thermal relaxation, and charge-phase heating may have a relatively small impact on discharge-phase temperatures. For duty cycles with shorter rest periods or multiple daily cycles, neglecting charge-phase heat generation may cause the model to underestimate mean cell temperatures and consequently underpredict degradation rates. Users modeling aggressive dispatch strategies should consider adjusting the ambient temperature offset or thermal resistance coefficient to account for this effect, or extend the framework to include explicit charge-phase thermal modeling.

### 3.8 Degradation Model

Total capacity loss is the sum of calendar and cycle contributions accumulated over the simulation horizon:

$$\text{SOH}(t) = 1 - Q_{\text{loss,cal}}(t) - Q_{\text{loss,cyc}}(t) \tag{9}$$

where SOH is bounded below by the configured end-of-life threshold $\text{SOH}_{\text{EOL}}$, at which point the asset is retired from service. Capacity loss from both calendar and cycle aging is accumulated at hourly resolution, with cell temperature evaluated as the midpoint average between consecutive timesteps. Within the discharge-only abstraction, degradation coefficients are treated as effective parameters that implicitly account for stress accumulated during the full operational cycle, including charging and rest periods not explicitly modeled. Details are provided in Supplementary Information S3.

*3.8.1 Calendar Aging*. Calendar aging is modeled as continuous capacity loss driven by temperature and state of charge, accumulating over elapsed time. The instantaneous calendar-aging rate follows a power-law dependence on elapsed time, modulated by temperature and state-of-charge stress factors:

$$\frac{dq_{\text{cal}}}{dt} = k_{\text{cal}} \, \beta \, t^{\beta-1} \, f_T(T) \, f_{\text{SOC}}(\text{SOC}) \tag{10}$$

where $q_{cal}$ is fractional capacity loss due to calendar aging, $k_{cal}$ is a rate constant (units: $hr^{-\beta}$), $\beta$ is the time exponent, and $f_T$ and $f_{SOC}$ represent temperature- and state-of-charge–dependent acceleration factors, respectively, as described below.

The time exponent $\beta$ governs the curvature of the degradation trajectory. Values of $\beta < 1$ produce sublinear degradation, while $\beta = 1$ corresponds to linear aging. Rather than fixing $\beta$ to a single value such as 0.5, SAGE treats it as a configurable parameter, allowing exploration of different calendar-aging behaviors under grid-relevant operating conditions. In the baseline configuration used for validation, $\beta = 0.75$, representing near-linear degradation with modest sublinearity. This choice is informed by field observations from utility-scale BESS installations[21, 26, 46], which consistently report approximately linear capacity fade over multi-year observation windows, in contrast to the roughly square-root kinetics sometimes[47] observed in controlled laboratory storage experiments.

Temperature acceleration follows an Arrhenius relationship:

$$f_T(T) = \exp\left[\frac{E_{a,cal}}{R}\left(\frac{1}{T_{ref}} - \frac{1}{T}\right)\right] \tag{11}$$

where $E_{a,cal}$ is the calendar-aging activation energy, $R$ is the gas constant, and $T_{ref}$ is a reference temperature. Elevated temperatures increase the instantaneous aging rate.

State-of-charge stress is captured through a multiplicative factor $f_{SOC}(SOC)$ that accelerates calendar aging at higher SOC, reflecting empirically observed behavior in lithium-ion cells[22, 48]

$$f_{SOC}(SOC) = \exp[\alpha_{cal} \cdot (SOC - SOC_{ref})] \tag{12}$$

where $\alpha_{cal}$ is the SOC stress coefficient and $SOC_{ref}$ is a reference state of charge.

Calendar aging accumulates continuously, including during idle periods, with the rate determined by the prevailing thermal and SOC conditions. The total calendar-induced capacity loss is obtained by numerically integrating the instantaneous rate over the simulation horizon.

*3.8.2 Cycle Aging.* Cycle aging accumulates proportionally to energy throughput[28] with Arrhenius temperature acceleration[22]:

$$\frac{dq_{cyc}}{dt} = k_{cyc} \cdot \frac{|P_{batt}(t)|}{E_{cap}} \cdot f_{T,cyc} \qquad (13)$$

where $q_{cyc}$ denotes the fractional capacity loss, $|P_{batt}|/E_{cap}$ represents instantaneous cycling rate in equivalent full cycles per hour, $k_{cyc}$ is capacity loss per equivalent full cycle at the reference temperature, and

$$f_{T,cyc} = \exp\left[\frac{E_{a,cyc}}{R}\left(\frac{1}{T_{ref}} - \frac{1}{T_{cell}}\right)\right] \qquad (14)$$

This formulation couples cycling damage to instantaneous cell temperature, capturing the physical reality that cycling at elevated temperatures accelerates degradation. Cycle aging accumulates only during active discharge; idle periods contribute zero cycle damage[49].

*3.8.3 Mechanism Independence.* The additive decomposition of capacity loss into calendar and cycle contributions assumes independent degradation pathways, an approximation that simplifies parameter identification while capturing dominant aging mechanisms[22]. This assumption is standard in semi-empirical lifetime models and is sufficient for isolating dominant stress drivers under grid-scale operating regimes. This formulation does not account for potential interactions between calendar and cycle aging, such as accelerated SEI reformation following cycling or synergistic effects at elevated temperatures[50]. The relative contributions of each mechanism emerge from simulation and depend on dispatch intensity, thermal environment, and configured rate constants.

### 3.9 Efficiency Fade

Discharge efficiency degrades linearly with state of health[8]:

$$\eta_{dis}(t) = \eta_{dis,BOL} - (\eta_{dis,BOL} - \eta_{dis,EOL}) \cdot \frac{1-\text{SOH}(t)}{1-\text{SOH}_{EOL}} \qquad (15)$$

This formulation creates an electro-thermal feedback loop: as internal resistance grows with aging, discharge efficiency declines, increasing resistive heat generation and thereby accelerating Arrhenius-driven degradation[5, 27]. This compounding behavior emerges from the coupled efficiency, thermal, and aging models rather than being imposed as an external constraint.

### 3.10 End-of-Life Criterion

Assets are retired when state of health falls below a fixed, configurable threshold[51], $\text{SOH}(t) \leq \text{SOH}_{\text{EOL}}$. The baseline threshold ($\text{SOH}_{\text{EOL}} = 0.70$) represents an effective operational end-of-life at which the combined effects of capacity loss, internal resistance growth, and thermal derating substantially reduce firm power capability, rendering grid-scale systems nonviable for primary energy and capacity services under typical market and safety constraints.

### 3.11 Fleet Simulation and Heterogeneity

Fleet simulations generate populations of assets sharing identical market environments and dispatch schedules but differing in intrinsic characteristics.

*3.11.1 Manufacturing Quality Factor*. Each asset receives a quality factor sampled from:

$$q_{\text{factor}} \sim \mathcal{N}(1.0, \sigma_{\text{qual}}) \tag{16}$$

Degradation rate constants are divided by this factor, representing cell-to-cell variability in effective reaction kinetics, interfacial stability, and defect density: values above 1.0 produce slower degradation (better cells) and values below 1.0 produce faster degradation (weaker cells). The baseline value ($\sigma_{\text{qual}} = 0.02$) corresponds to 2% coefficient of variation, consistent with reported manufacturing tolerances for commercial cells[21, 52].

*3.11.2 Rack Position*. Each asset receives a vertical position sampled uniformly:

$$g \sim \mathcal{U}(0,1) \tag{17}$$

This parameter determines the static thermal offset, capturing persistent enclosure-scale thermal stratification documented in field studies rather than transient airflow effects[26].

### 3.12 Measurement Model

SAGE exports both ground-truth states and sensor-corrupted measurements:

$$\text{SOC}_{\text{meas}}(t) = \text{clip}(\text{SOC}_{\text{true}}(t) + \varepsilon_{\text{SOC}}, 0, 1) \tag{18}$$

$$\text{SOH}_{\text{meas}}(t) = \text{clip}(\text{SOH}_{\text{true}}(t) + \varepsilon_{\text{SOH}}, 0, 1) \tag{19}$$

$$T_{\text{cell,meas}}(t) = T_{\text{cell,true}}(t) + \varepsilon_T \tag{20}$$

where $\varepsilon_{\text{SOC}} \sim \mathcal{N}(0, \sigma_{\text{SOC}}^2)$, $\varepsilon_{\text{SOH}} \sim \mathcal{N}(0, \sigma_{\text{SOH}}^2)$, and $\varepsilon_T \sim \mathcal{N}(0, \sigma_T^2)$. SOC and SOH are clipped to [0, 1] because values outside this range are physically meaningless; temperature requires no clipping. Baseline noise levels ($\sigma_{\text{SOC}} = 0.02$, $\sigma_{\text{SOH}} = 0.01$, $\sigma_T = 0.5°C$) reflect estimated commercial BMS measurement uncertainty[4, 52, 53]. This dual-output design enables training on realistic noisy data while validating against known ground truth.

### 3.13 Sensitivity Analysis Framework

SAGE includes an automated sensitivity analysis framework designed to quantify how uncertainty in thermal, operational, and degradation parameters propagates into fleet-level lifespan outcomes. Sensitivity studies are performed using a one-at-a-time (OAT) methodology, in which a single parameter is perturbed about a common baseline while all other parameters are held fixed. This approach isolates direct causal effects, preserves physical interpretability, and avoids confounding interactions that can obscure mechanistic attribution in long-horizon simulations.

The primary outputs of interest are fleet-average lifespan and distributional metrics (e.g., dispersion and tail behavior), rather than individual asset trajectories. This focus reflects the intended use of SAGE as a population-level analysis and benchmarking tool, where tail risk and heterogeneity are often more consequential than representative mean behavior[6].

*3.13.1. Parameter Classes and Physical Roles*. Parameters selected for sensitivity analysis span three distinct physical roles:

1. Thermal boundary conditions and operational stressors, including the HVAC-controlled container baseline temperature and the daily discharge duration. These parameters establish persistent thermal offsets and cumulative cycling stress that compound over the full asset lifetime.
2. Electrochemical degradation kinetics, including the activation energies governing calendar and cycle aging. These parameters control the temperature sensitivity and stress weighting of degradation accumulation within the Arrhenius-based aging models.
3. Intrinsic heterogeneity, represented by manufacturing quality variability, which introduces asset-to-asset dispersion without shifting the fleet mean by construction.

These parameters were chosen because they either define long-lived thermal environments, set cumulative operational stress, or introduce persistent heterogeneity across otherwise identical operating histories.

*3.13.2. Sensitivity Quantification.* For each parameter, fleet simulations are repeated across a predefined sweep range centered on the baseline configuration. All sensitivity runs use fixed random seeds so that observed differences in lifespan statistics reflect parameter effects rather than Monte Carlo noise. Sensitivities are quantified using both absolute changes in mean fleet lifespan and local elasticities, defined as the percent change in mean lifespan per percent change in parameter value evaluated about the baseline.

*3.13.3. Methodological Scope and Limitations.* Although higher-order interactions between parameters may exist—particularly between temperature and cycling stress—the OAT framework is sufficient for identifying first-order sensitivities in a physics-informed model where parameters have clear mechanistic meaning. Interaction effects are intentionally deferred to future work, as their interpretation depends strongly on site-specific operating regimes, dispatch strategies, and thermal management designs.

A complete specification of baseline values, sweep ranges, and step sizes used in the sensitivity studies is provided in Supplementary Table S3.

Although the baseline dispatch policy is intentionally simplified to enable controlled validation, the mechanisms generating degradation heterogeneity arise from asset-specific thermal exposure and stochastic environmental variation rather than from dispatch optimality, and therefore remain applicable across alternative operating strategies. Because the dispatch policy is designed to generate realistic stress exposure rather than revenue-optimal operation, economic outcomes should be interpreted qualitatively, with emphasis placed on lifecycle stress patterns and degradation trajectory evolution

### 3.14 Simulation Parameters

Table 1 summarizes the main baseline parameter values; complete parameter specifications are provided in Supplementary Information Table S1. All parameters are configurable; values shown represent a system with 5 MWh nameplate capacity and 1 MW discharge power in a nominally ERCOT-like market. This modest energy margin is used here as a baseline modeling choice to avoid immediate energy-limited operation at beginning of life and to isolate aging-driven transitions under otherwise fixed dispatch. The margin is not intended to represent an optimal design choice; SAGE is fully configurable and the energy-to-power ratio can be varied systematically.

### 4. RESULTS AND DISCUSSION

This section evaluates SAGE's internal physical consistency through hierarchical validation checks spanning the market environment, dispatch behavior, single-asset degradation physics[31], and fleet-level statistics. Validation is performed against established theoretical expectations and emergent statistical properties of electrochemical degradation and electricity markets, rather than calibration to proprietary field datasets that remain largely inaccessible. Metrics reported here are extracted from simulation outputs rather than imposed as constraints, providing a stringent test of correct physics implementation. The results were obtained using fleet_size (number of modules in the fleet) = 1,000.

## 4.1 Market Environment Calibration

To support long-horizon degradation analysis, the synthetic market environment must provide economically plausible price levels and include rare scarcity events that materially influence dispatch and thermal stress, without attempting detailed market forecasting or historical replication. Figure 1 provides a minimal validation of these two requirements.

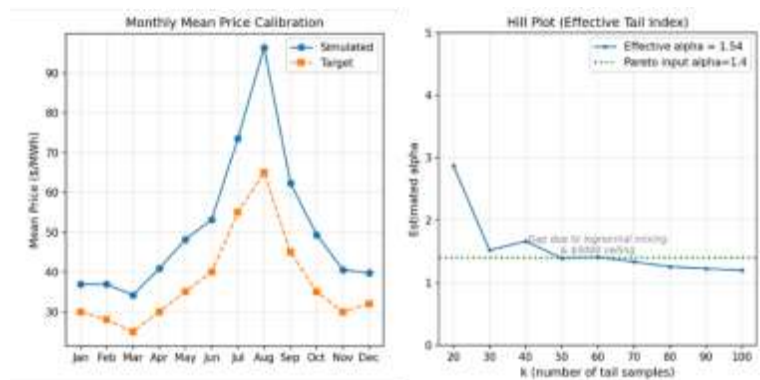

Figure 1(a) compares monthly mean electricity prices produced by the synthetic generator with a reference seasonal profile. The synthetic prices exhibit the correct seasonal structure, including elevated summer pricing and lower winter values, and fall within a realistic magnitude range for an ERCOT-like energy-only market. Summer means exceed the reference profile due to explicit weather-driven demand excursions and temperature-dependent shifts in the timing and width of the daily price peak, reflecting the stochastic stress environment rather than calibration to historical averages. Although closer agreement could be achieved through parameter adjustment, SAGE is not intended as a curve-fitting tool. Its purpose is to produce structurally realistic, stress-inducing price environments suitable for controlled benchmarking, rather than to replicate a specific historical dataset.

Figure 1(b) estimates the effective tail exponent using a Hill plot. The Hill plot estimates how "heavy" the extreme-price tail is by fitting a Pareto-like (power law) exponent using only the largest $k$ price events; a flat region suggests a reliable tail exponent. As the number of upper-tail samples $k$ increases, the estimated exponent $\hat{\alpha}(k)$ approaches a stable plateau near the imposed Pareto value, indicating that the upper tail is consistent with power-law (scarcity-driven) behavior over an intermediate range of $k$. Deviations at very small $k$ reflect finite-sample variance in the Hill estimator. At larger $k$, systematic deviations arise from mixing between scarcity-driven Pareto tails and the non-Pareto price backbone, as well as truncation imposed by the market price cap. Details are given in the Supplementary Information, section S2.2.3.

Together, these validations establish that the synthetic environment provides realistic economic forcing and scarcity stress suitable for degradation-focused simulation, without embedding application-specific market optimization or forecasting assumptions.

### 4.2 Dispatch Behavior and Electro-Thermal Coupling

We next examine how price-driven dispatch decisions interact with the synthetic market environment and how resulting operating commands propagate into electro-thermal response at the asset level. The purpose of this analysis is not to demonstrate optimality or forecasting performance, but to establish that dispatch behavior emerges coherently from the stochastic price signal and that electrical loading produces physically consistent thermal stress.

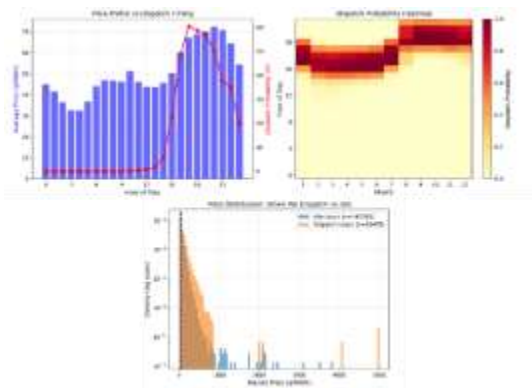

Figure 2a illustrates the imposed daily discharge structure used throughout this study. Dispatch is restricted to a fixed window (11:00–21:00) and a single contiguous block per day, reflecting common grid-scale operating constraints rather than an optimized control policy. The purpose of this figure is to provide context for subsequent electro-thermal and degradation dynamics, not to demonstrate dispatch optimality.

Seasonal and diurnal structure in dispatch behavior is further illustrated in the dispatch probability heatmap shown in Figure 2(b). Elevated dispatch probabilities persist across all months during late-day peak hours, while mid-day shoulder periods exhibit weaker seasonal modulation. Morning and overnight hours remain largely inactive throughout the year. The consistency of this pattern across seasons indicates that dispatch timing emerges from the stochastic market environment rather than from deterministic scheduling assumptions. Minor seasonal shifts in the timing of peak dispatch are observed, reflecting temperature-driven shifts in the daily price profile rather than fixed scheduling assumptions.

Figure 2(c) compares market prices conditional on dispatch and idle states on a semi-log graph. Prices observed during dispatch hours are systematically shifted toward higher values relative to idle periods, with a greater incidence of extreme prices in the upper tail. This conditional separation confirms that the dispatch logic preferentially samples economically favorable operating windows rather than uniformly cycling across the price distribution.

Together, these results demonstrate that SAGE links stochastic market conditions, price-responsive dispatch behavior, and electro-thermal physics within a unified simulation framework. Dispatch activity adapts naturally to the statistical structure of the synthetic price signal, while thermal states evolve coherently in response to operating stress. This establishes a physically meaningful foundation for subsequent degradation and fleet-level analyses.

**4.3 Degradation Trajectories and Mechanism Decomposition**

Having established realistic market environments and price-responsive dispatch behavior, we next examine whether individual assets exhibit degradation trajectories consistent with the configured electrochemical aging mechanisms. The objective here is not empirical lifetime prediction, but to demonstrate internally consistent accumulation of calendar- and cycle-driven capacity loss under representative operating conditions.

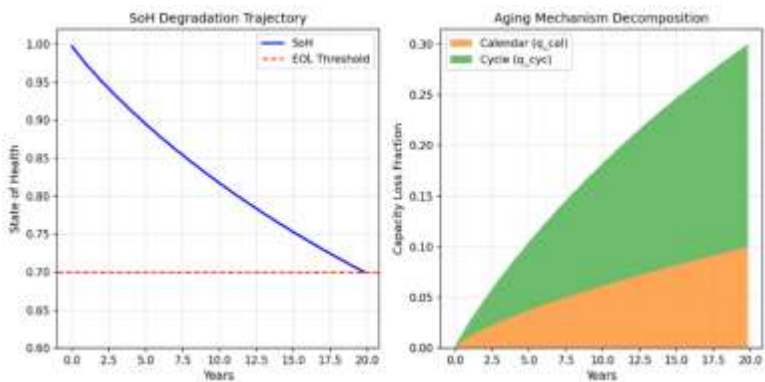

Figure 3(a) shows the state-of-health trajectory for a representative asset operating under daily arbitrage dispatch. The trajectory exhibits the smooth, concave shape expected from sublinear calendar aging kinetics combined with cumulative cycling effects, reaching the configured end-of-life threshold after approximately 20 years under baseline conditions. This behavior reflects the prescribed time dependence and confirms stable numerical integration of degradation rates over long horizons.

Figure 3(b) decomposes cumulative capacity loss into calendar and cycle contributions. Both mechanisms contribute from early operation, with cycle aging accumulating approximately linearly with cumulative throughput and calendar aging following the imposed sublinear dependence on elapsed time. By end of life, cycle aging accounts for the dominant fraction of total capacity loss under the assumed single daily discharge profile, consistent with moderate cycling intensity.

Together, these trajectories demonstrate that SAGE propagates degradation mechanisms coherently over time and preserves their intended physical roles, providing a mechanistically interpretable foundation for subsequent fleet-level and sensitivity analyses.

## 4.4 Thermal–Degradation Coupling

Temperature exerts a dominant influence on battery degradation kinetics through Arrhenius scaling. This section evaluates whether SAGE's coupled thermal–degradation implementation produces physically consistent temperature dependence by extracting effective activation energies directly from full fleet simulations and comparing them with the configured model parameters[5].

The thermal environment introduces controlled heterogeneity across the simulated fleet. Assets located at higher vertical positions within the rack experience systematically elevated cell temperatures due to persistent thermal stratification, with a configured gradient of 5 °C from bottom to top. This gradient remains stable over the simulation horizon and produces an ensemble of otherwise identical assets that differ only in their thermal histories, providing a natural basis for Arrhenius analysis without introducing artificial parameter variation.

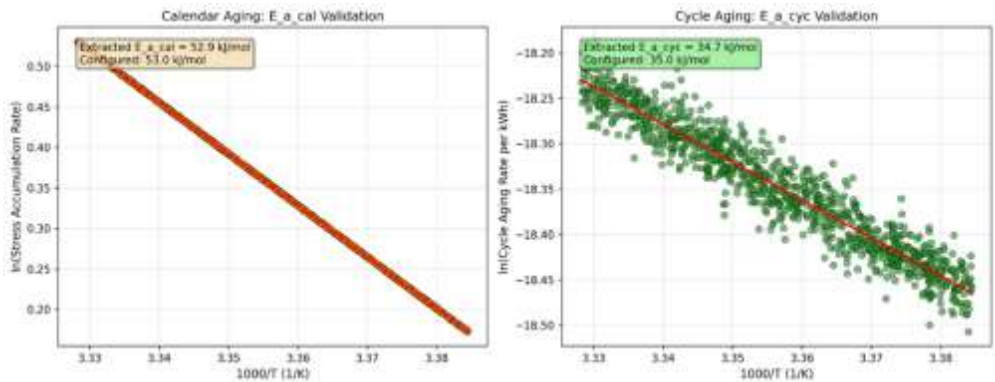

Figure 4 presents an Arrhenius consistency check for the calendar- and cycle-aging implementations. For calendar aging, rather than analyzing raw capacity fade rates, we construct a normalized stress-accumulation metric based on an effective aging time that weights elapsed time by the instantaneous temperature and state-of-charge stress. Define the effective aging time for calendar aging as

$$t_{\text{eff}}(t) = \int_0^t \underbrace{A_{\text{cal}}(T(\tau))}_{\text{Arrhenius factor}} \underbrace{f_{\text{SOC}}(\text{SOC}(\tau))}_{\text{SOC stress factor}} d\tau$$

where a standard Arrhenius acceleration factor (normalized to 1 at $T_{\text{ref}}$) is

$$A_{\text{cal}}(T) = \exp\left[\frac{E_{a,\text{cal}}}{R}\left(\frac{1}{T_{\text{ref}}} - \frac{1}{T}\right)\right]$$

$f_{\text{SOC}}(\text{SOC})$ is the SOC-dependent stress function defined in Eq. (12). We then define the normalized stress-accumulation rate as $t_{\text{eff}}/t$, representing the ratio of Arrhenius- and SOC-weighted effective aging time to elapsed clock time. By construction, this metric removes the calendar-aging rate constant and isolates the temperature-dependent acceleration by evaluating the Arrhenius- and SOC-weighted stress independently of the intrinsic time-power kernel of the calendar-aging formulation. Consequently, asset-to-asset variability in absolute aging rate does not influence this analysis.

When the natural logarithm of the normalized (time-averaged) stress-accumulation rate is plotted against inverse absolute temperature, the data collapse onto a near-perfect linear relationship (Fig. 4a). Linear regression yields an effective activation energy $E_{a,cal}$ = 52.9 kJ/mol, in close agreement with the configured input value of 53.0 kJ/mol, confirming correct propagation of Arrhenius temperature dependence in the calendar-aging pathway.

For cycle aging, we analyze a throughput-normalized capacity-loss metric rather than an Arrhenius-weighted stress time, corresponding to degradation per unit energy throughput. In contrast to the calendar-aging construction, this quantity intentionally retains the cycle-aging rate constant and therefore inherits asset-to-asset variability arising from manufacturing quality differences. In addition, cycle aging accumulates only during intermittent dispatch events rather than continuously in time, resulting in a substantially smaller set of Arrhenius-relevant observations. As a consequence, the Arrhenius plot exhibits measurable scatter about the fitted trend (Fig. 4b). Despite this dispersion, plotting the natural logarithm of throughput-normalized capacity loss against inverse absolute temperature yields a clear linear relationship. Linear regression extracts an effective activation energy $E_{a,cyc}$ = 34.7 kJ/mol, in close agreement with the configured input value of 35 kJ/mol, with scatter affecting the intercept but not the Arrhenius slope, confirming correct propagation of Arrhenius temperature dependence in the cycle-aging pathway.

The contrast between panels (a) and (b) is therefore intentional: the calendar-aging construction isolates temperature acceleration alone, while the cycle-aging analysis retains manufacturing heterogeneity and dispatch intermittency, causing physically meaningful dispersion without altering the Arrhenius slope. These Arrhenius relationships emerge from full dynamic simulations rather than from direct parameter insertion, providing a strong internal consistency check on the thermal model, degradation formulations, and their numerical integration.

**4.5 Thermal Stratification**

To validate the implementation of thermal stratification and its propagation into degradation outcomes, we examine the relationship between rack position and operating temperature.

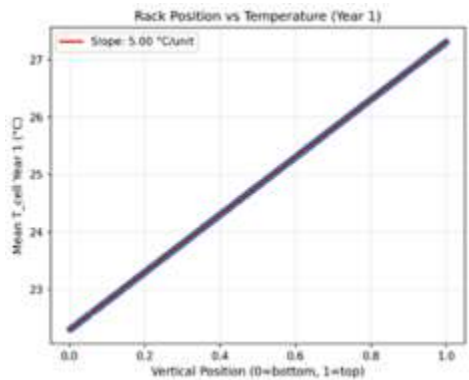

Figure 5 shows mean cell temperature during the first year of operation as a function of vertical rack position. Temperature varies linearly with position, with a fitted slope of 5.0

°C per unit height, in exact agreement with the configured vertical thermal gradient. The near-perfect linearity reflects the structure of the thermal model: all assets experience identical ambient conditions and nearly identical operating behavior (SOH ≈ 1), such that rack position is the dominant source of temperature variation. This panel therefore provides a direct validation of the deterministic mapping from physical location to cell temperature.

### 4.6 Comparison to Simplified Degradation Models

To quantify the implications of physics-based degradation modeling, SAGE predictions are compared against two simplified baseline models representative of approaches frequently used in practice (i) a linear degradation model assuming constant state-of-health (SOH) decline over time, typical of financial pro forma analyses; and (ii) a throughput-only model, in which degradation scales linearly with cumulative equivalent full cycles, commonly used in warranty structures and cycle-life specifications. These models are intentionally minimal and do not represent thermal feedback, nonlinear kinetics, or asset-level heterogeneity.

The linear model assumes a constant fade rate over calendar time:

$$\text{SOH}(t) = 1 - \frac{1 - \text{SOH}_{\text{EOL}}}{t_{\text{EOL}}} \cdot t \tag{21}$$

where $t_{\text{EOL}}$ is the end-of-life time at which the asset reaches the threshold $\text{SOH}_{\text{EOL}}$.

The throughput-only model assumes degradation proportional to cumulative energy throughput:

$$\text{SOH}(E) = 1 - \frac{1 - \text{SOH}_{\text{EOL}}}{E_{\text{EOL}}} \cdot E_{\text{cumulative}} \tag{22}$$

where $E_{\text{EOL}}$ is the cumulative throughput corresponding to end-of-life and $E_{\text{cumulative}}$ is the accumulated equivalent full-cycle throughput. Both baseline models are calibrated to match the SAGE end-of-life time for a reference asset to ensure that comparisons reflect differences in trajectory shape and structural behavior rather than absolute lifetime scaling.

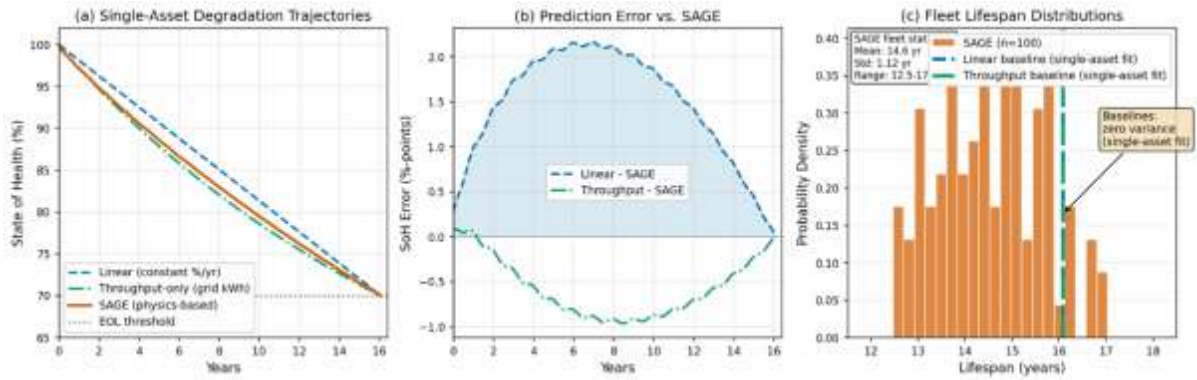

Figure 6(a) compares single-asset SOH trajectories predicted by the two baseline models and by SAGE under identical dispatch conditions. The linear model predicts constant degradation throughout life and fails to reproduce the concave trajectory arising from sublinear calendar-aging kinetics in SAGE. Under the specific operating conditions examined here, the throughput-only model closely tracks the SAGE trajectory because cumulative throughput accumulates nearly linearly in time for the fixed daily discharge schedule. Error analysis relative to SAGE (Fig. 6(b)) highlights these structural differences: the linear model exhibits systematic mid-life error due to its inability to represent nonlinear aging kinetics, whereas the throughput-only model shows lower error under this regular dispatch regime.

The most significant distinction emerges at fleet scale (Fig. 6(c)). When applied to a heterogeneous ensemble with variability in thermal environment and manufacturing quality, SAGE produces a finite distribution of asset lifetimes reflecting emergent physical heterogeneity. In contrast, the reduced-order baseline models predict identical lifetimes for all assets because they contain no mechanism for representing asset-level variability. This comparison highlights a fundamental limitation of simplified degradation models: even when calibrated to match single-asset end-of-life behavior, they cannot reproduce the population-level dispersion that emerges naturally from coupled thermal and degradation physics.

## 4.7 Sensitivity of Fleet-Average Lifespan to Physical and Operational Parameters

To identify which physical and operational uncertainties most strongly control fleet-average lifespan, we performed a one-at-a-time (OAT) sensitivity analysis in which individual model inputs were perturbed about a common baseline while all other parameters were held fixed. Parameters examined here are restricted to thermal boundary conditions, operational stressors, degradation kinetics, and intrinsic heterogeneity. To enable comparison across parameters with different physical units, sensitivities are reported as local elasticities, defined as the percent change in mean fleet lifespan per percent change in parameter value, evaluated about the baseline configuration. The sensitivity calculations were run with fleet_size = 100.

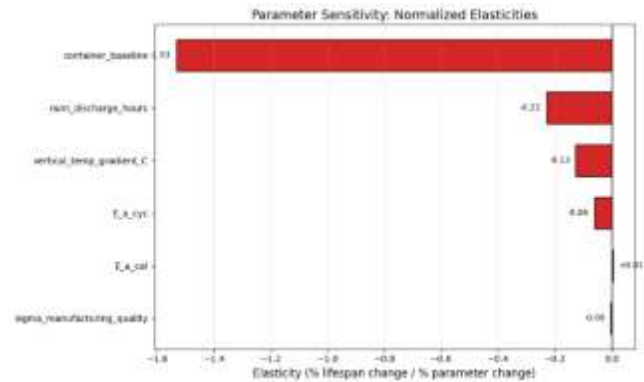

Figure 7 summarizes the resulting elasticities. Negative elasticities indicate that increasing the parameter reduces fleet-average lifespan, while positive values indicate lifespan extension.

*4.7.1 Dominance of Mean Thermal Stress*. Persistent thermal boundary conditions dominate fleet-average lifespan. The container baseline operating temperature, representing the effective mean temperature experienced by the cells, exhibits by far the largest elasticity ($\epsilon \approx -1.53$). This parameter directly sets the thermal boundary condition entering Arrhenius acceleration of both calendar and cycle aging. The magnitude of this elasticity implies that a sustained 10% increase in mean operating temperature (approximately 2 °C about the baseline) reduces mean fleet lifespan by roughly 16 %.

This result highlights a fundamental separation between economic and physical drivers of degradation. While outdoor climate strongly influences electricity prices and dispatch incentives, it affects aging primarily through its impact on the internal thermal environment of the cells. When thermal management is effective, longevity is governed by container operating temperature rather than by external weather conditions. Thermal control therefore acts as the primary lever by which operators decouple asset lifetime from site-specific climate variability.

*4.7.2 Role of Operational Intensity.* Operational intensity, represented by the number of daily discharge hours, is the second most influential parameter ($\epsilon \approx -0.23$). Increasing discharge duration directly raises cumulative throughput and elevates thermal stress during dispatch windows, accelerating both cycle-driven degradation and temperature-dependent aging mechanisms.

This result underscores that how hard an asset is used is significant, not just how hot it is kept. Even under well-controlled thermal conditions, more aggressive utilization materially shortens fleet-average lifetime. Operational policy therefore represents a distinct and powerful lever for lifetime management, complementary to thermal design and HVAC control.

*4.7.3 Thermal Stratification and Spatial Heterogeneity.* The vertical thermal gradient within the enclosure exhibits a smaller but still meaningful elasticity ($\epsilon \approx -0.13$). Unlike the mean operating temperature, which shifts the degradation trajectory of the entire fleet uniformly, thermal stratification introduces spatially localized aging. Modules located in persistently hotter regions age more rapidly and reach end of life earlier, creating a weakest-link constraint on fleet operation.

This distinction has important practical implications. Lowering the global operating temperature to compensate for a large thermal gradient requires over-cooling the coolest modules, increasing auxiliary power consumption without eliminating the underlying non-uniformity. The sensitivity results therefore separate operational levers (global temperature setpoints and dispatch intensity) from design constraints (airflow uniformity and enclosure geometry). Mitigating stratification requires improvements in thermal design rather than simply more aggressive cooling.

*4.7.4 Secondary Influence of Degradation Kinetics*. Uncertainty in degradation kinetics plays a substantially smaller role under the baseline operating regime considered here. The cycle-aging activation energy exhibits a modest negative elasticity ($\epsilon \approx -0.06$), while the calendar-aging activation energy shows near-zero sensitivity. This ordering is physically consistent: the leverage of Arrhenius kinetics depends on the magnitude of temperature deviations from the reference condition, and effective thermal management limits such deviations over most of the operating life.

These sensitivities should be interpreted as regime-dependent rather than intrinsic insensitivity of the underlying kinetics. Because of the non-linearity of the Arrhenius expression, at higher baseline operating temperatures Arrhenius acceleration would be substantially stronger, and uncertainties in activation energies would exert a correspondingly larger influence on lifetime outcomes.

*4.7.5 Manufacturing Variability*. Manufacturing quality variability exhibits negligible elasticity with respect to mean fleet lifespan ($\epsilon \approx 0$). This behavior is expected, as variability is implemented symmetrically about unity and therefore does not shift the fleet average. Its primary effect is to increase dispersion and early-failure risk rather than to alter the central tendency. Parameters governing heterogeneity and tail behavior therefore influence reliability and warranty exposure rather than mean lifespan metrics.

*4.7.6 Summary Interpretation*. Figure 7 demonstrates that fleet-average aging behavior is dominated by thermal management and operational intensity, acting through distinct but complementary mechanisms. Mean operating temperature sets the global aging rate of the fleet, while dispatch duration controls cumulative stress imposed on the assets. Thermal stratification introduces additional lifetime penalties through spatial heterogeneity, whereas uncertainties in degradation kinetics and manufacturing variability play secondary roles under effective thermal control at the assumed temperature.

Together, these results establish thermal design, HVAC operation, and dispatch policy—not electrochemical parameter uncertainty—as the primary determinants of long-term BESS longevity at fleet scale under our assumed operating conditions. Full numerical results and parameter sweep ranges are provided in the Supplementary Information.

**4.8 Limitations**

Several limitations of the present implementation should be considered when interpreting the results.

*4.8.1 Degradation mechanisms*. The model considers capacity fade driven by SEI growth and cycling stress. Additional mechanisms, such as lithium plating, electrode cracking, and electrolyte decomposition, are not included and could require model extensions for applications involving fast charging or extreme temperatures.

*4.8.2 Thermal dynamics.* A quasi-steady-state thermal approximation is employed, assuming cell temperature responds instantaneously to operating conditions. This approximation is appropriate for the baseline dispatch regime (one 4-hour block per day at approximately C/4), where discharge duration substantially exceeds typical enclosure thermal time constants and temperatures approach steady-state within each hour. For high-power applications such as frequency response or fast-ramping ancillary services, where C-rates may reach 1C or higher and discharge durations are measured in minutes rather than hours, transient thermal spikes would exceed steady-state predictions. The quasi-steady-state assumption would therefore underestimate peak cell temperatures and consequently underpredict degradation rates for such applications. Extending SAGE to high-power duty cycles could require a thermal mass model to capture transient heat retention between cycling events.

*4.8.3 Charging physics.* The current version of SAGE employs a discharge-only abstraction where degradation coefficients are treated as effective, full-cycle aggregates. This approach assumes that the stress accumulated during the charging and rest phases is implicitly captured within the discharge-phase physics. While this simplifies the model for energy arbitrage scenarios, it introduces several limitations for high-fidelity research:

- **Thermal Underestimation**: Neglecting heat generation during the charge phase prevents the model from capturing the cumulative thermal state of the battery. In real systems, charging elevates the cell temperature prior to the onset of discharge, meaning the quasi-steady-state approximation may underestimate the peak cell temperatures and the resulting Arrhenius-driven degradation rates.
- **State-of-Charge Stress**: Calendar aging is highly sensitive to the time an asset spends at high SOC. By treating charging as an exogenous boundary condition, the framework does not explicitly resolve how different charging durations and protocols—such as slow trickle charging versus aggressive fast charging—alter the "time-at-voltage" stress.
- **Parameter Absorption**: Under the baseline dispatch of one cycle per day, these effects are largely absorbed into the "effective" degradation rates. This abstraction preserves physical consistency at the cycle level while avoiding asset-specific charging assumptions that vary fundamentally across different system co-locations.
- **Economic-Physical Coupling**: Although SAGE is not an economic engine, the absence of grid import costs limits its utility as a benchmarking platform for "degradation-aware" optimization. True optimization algorithms must balance the market cost of electricity (the "buy" price) against the physical degradation cost incurred during that specific charging event.
- **Asset-Specific Realism**: Applications requiring explicit representation of site-specific charging protocols or grid import costs would require framework extensions to resolve these sub-hourly dynamics. These limitations will be addressed in upcoming work to further enhance the framework's fidelity.

4.*8.4 Dispatch flexibility.* The degradation physics (Arrhenius-weighted calendar and cycle aging) are formulated independently of dispatch strategy. However, the current simulation loop assumes daily full recharge and block-committed power levels. Adapting SAGE to irregular or externally generated power profiles would require modifying these operational assumptions while retaining the core degradation equations.

*4.8.5 Field validation.* Comprehensive quantitative validation against proprietary fleet datasets remains future work. The qualitative comparisons in Section 4.7 support physical plausibility but do not constitute calibration to specific fielded assets.

## 5. CONCLUSIONS

We have presented SAGE, an open-source, physics-informed simulation framework for generating hourly resolved, long-horizon synthetic operating histories and degradation trajectories for heterogeneous battery energy storage system fleets under realistic dispatch. Through hierarchical validation spanning stochastic environment generation, single-asset degradation kinetics, electro-thermal coupling, and fleet-level statistics, we demonstrate that SAGE produces internally consistent and physically interpretable behavior suitable for applications that are currently constrained by the scarcity of long-duration field datasets.

The framework reproduces key structural features of grid operating environments, including seasonal variability and heavy-tailed scarcity events, enabling realistic stress patterns for battery operation. At the asset level, degradation trajectories follow the configured electrochemical kinetics, with temperature-dependent aging behavior consistent with Arrhenius theory and physically consistent coupling between efficiency fade, heat generation, and state evolution. Thermal stratification and manufacturing variability across assets produce systematic differences in operating temperature and degradation rate, providing a controlled mechanism for studying heterogeneous aging behavior. This behavior arises naturally from coupled physics and stochastic variability without imposing statistical failure models, enabling reproducible investigation of population-level degradation dynamics that are not accessible in representative-asset or throughput-only modeling approaches.

These results establish SAGE as a flexible platform for generating physically grounded synthetic datasets to support development and benchmarking of degradation-aware algorithms, state-estimation methods, machine-learning workflows, and fleet-scale sensitivity analyses under controlled conditions. By enabling systematic exploration of parameter uncertainty and heterogeneity using transparent, configurable models, the framework supports reproducible research in grid-scale energy storage modeling without requiring decades of field operation. The present implementation intentionally abstracts selected electrochemical and operational complexities to maintain interpretability and controlled validation, while the modular structure enables straightforward incorporation of these effects in future extensions.

Stephen J. Harris: Conceptualization, Methodology, Software, Validation, Formal Analysis, Writing – Original Draft, Writing – Review & Editing. Paolo Esquivel: Conceptualization, Software. Mark A. Harris: Conceptualization

The authors declare that they have no known competing financial interests or personal relationships that could have appeared to influence the work reported in this paper.

SAGE is released as open-source software and is available at https://github.com/sagebessbatteries/sage, along with example configurations, documentation, and validation scripts.

# Table 1

| Category | Parameter | Symbol | Value | Units | Description |
|---|---|---|---|---|---|
| System | Nominal energy capacity | $E\_nom$ | 5 | MWh | Nameplate energy capacity at beginning of life |
| System | Nominal discharge power | $P\_nom$ | 1 | MW | Nameplate discharge power |
| System | Discharge duration | H | 4 | h | Fixed-duration daily discharge block |
| Thermal | HVAC temperature setpoint | $T\_set$ | 22 | °C | Target container air temperature |
| Thermal | Vertical thermal gradient | $\Delta T\_grad$ | 5 | °C | Bottom-to-top rack temperature difference |
| Operation | Daily discharge frequency | — | 1 | cycle/day | At most one discharge block per day |
| SOC limits | Initial SOC window | $SOC\_min$–$SOC\_max$ | 0.05–0.95 | — | Beginning-of-life usable SOC range |
| SOC limits | End-of-life SOC window | $SOC\_min$–$SOC\_max$ | 0.20–0.80 | — | End-of-life usable SOC range |
| Lifetime | End-of-life threshold | $SoH\_EOL$ | 0.7 | — | Operational retirement criterion |
| Market | Market archetype | — | ERCOT-like | — | Synthetic energy-only market with scarcity pricing |

## Table 2

| Category | Parameter | Symbol | Value | Units | Description |
|---|---|---|---|---|---|
| Calendar aging | Calendar aging rate constant | k_cal | 1.00E-05 | 1/h | Base calendar aging rate at reference conditions |
| Calendar aging | Calendar aging time exponent | beta_cal | 0.75 | — | Time dependence of calendar aging |
| Calendar aging | SOC stress coefficient | alpha_cal | 1.5 | — | Scaling coefficient for SOC-dependent calendar aging |
| Calendar aging | Calendar aging activation energy | E_a,cal | 53000 | J/mol | Arrhenius activation energy for calendar aging |
| Cycle aging | Cycle aging rate constant | k_cyc | 5.00E-05 | 1/cycle | Base cycle aging rate at reference temperature |
| Cycle aging | Cycle aging activation energy | E_a,cyc | 42000 | J/mol | Arrhenius activation energy for cycle aging |
| Thermal | Container temperature setpoint | T_set | 22 | °C | HVAC-regulated container air temperature |
| Thermal | Container attenuation factor | α_cont | 0.0833 | — | Fraction of outdoor temperature swing transmitted indoors |
| Thermal | Operating temperature rise at C/4 | ΔT_C/4 | 2 | °C | Steady-state temperature rise during C/4 discharge |
| Thermal | Vertical rack temperature gradient | ΔT_grad | 5 | °C | Bottom-to-top temperature difference within rack |
| Thermal | Maximum cell temperature | T_cell,max | 55 | °C | Thermal safety limit for cell operation |
| Thermal | Arrhenius reference temperature | T_ref | 298.15 | K | Reference temperature for Arrhenius scaling |
| Efficiency | Discharge efficiency at BOL | η_BOL | 0.95 | — | Beginning-of-life discharge efficiency |
| Efficiency | Discharge efficiency at EOL | η_EOL | 0.9 | — | End-of-life discharge efficiency |
| Heterogeneity | Manufacturing quality std dev | σ_qual | 0.02 | — | Standard deviation of asset quality factor |
| Heterogeneity | Fleet size | N | 100 | assets | Number of simulated battery modules |
| Measurement | SOC measurement noise | σ_SOC | 0.02 | — | Standard deviation of SOC measurement error |
| Measurement | SoH measurement noise | σ_SoH | 0.01 | — | Standard deviation of SoH measurement error |
| Measurement | Temperature measurement noise | σ_T | 0.5 | °C | Standard deviation of temperature sensor noise |

**FIGURES AND CAPTIONS**

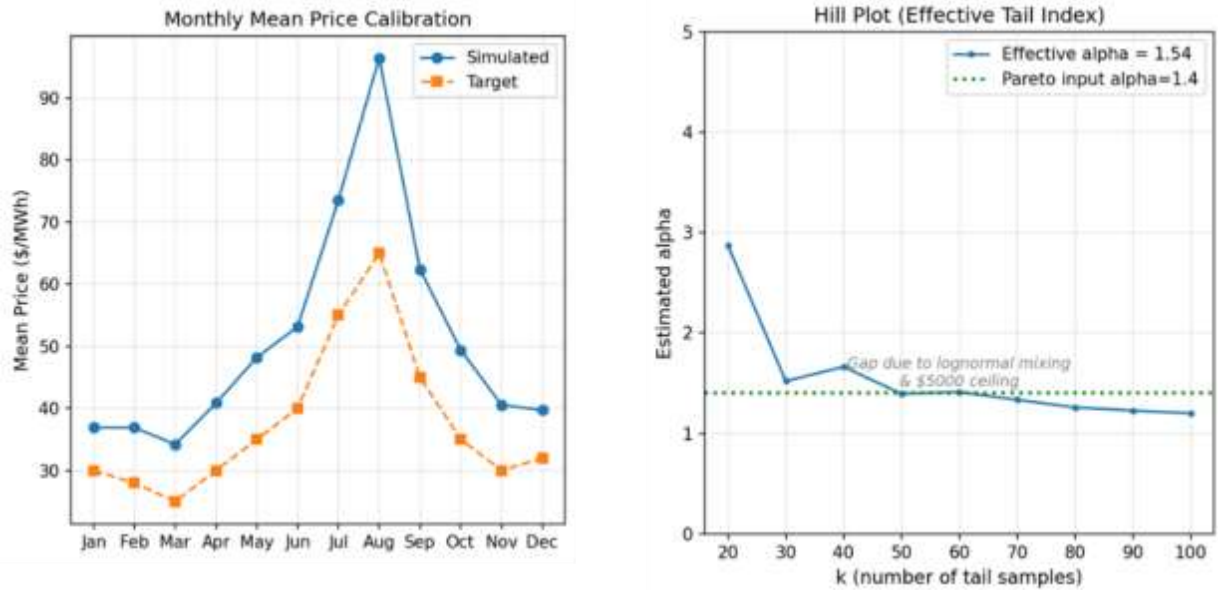

**Figure 1. Validation of the synthetic market environment.** (a) Monthly mean electricity prices for the reference seasonal profile and the synthetic generator, demonstrating realistic price levels and correct seasonal structure. The synthetic environment includes weather-driven excursions around the reference profile to represent stochastic operating stress rather than historical calibration. (b) Log–log complementary cumulative distribution function (CCDF) of electricity prices, showing heavy-tailed scarcity behavior. The fitted tail exponent closely matches the imposed Pareto parameter, confirming that extreme price events arise naturally from the stochastic process.

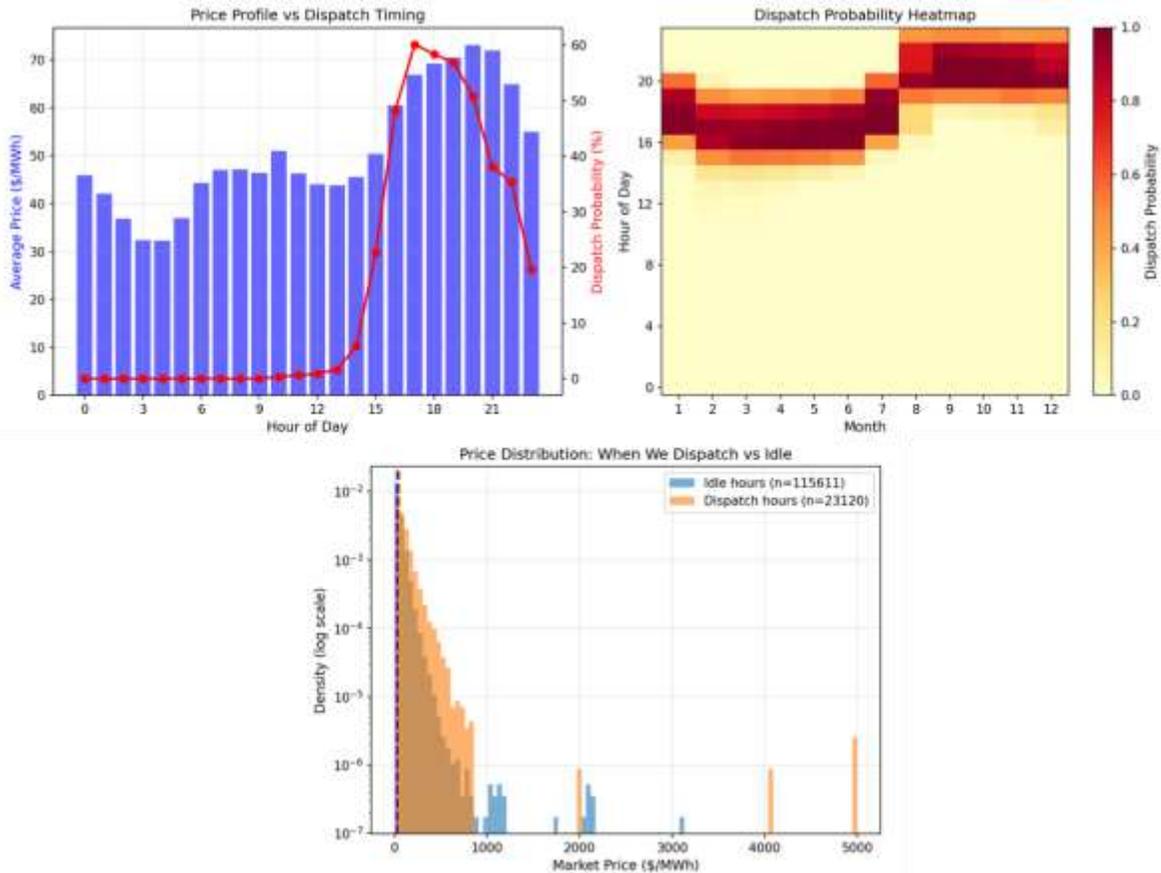

**Figure 2. Imposed dispatch structure and resulting electro-thermal operating context.** (a) Average hourly electricity prices (bars) and empirical probability of dispatch (line) as a function of hour of day. Dispatch is restricted by design to a fixed daily window (11:00–21:00) and a single contiguous block per day; the concentration of dispatch activity during higher-price hours reflects this imposed operating abstraction together with learned scheduling. (b) Dispatch probability as a function of hour of day and month, illustrating the consistent late-day operating window used throughout the simulations across seasons. This structure provides a controlled and repeatable operating context for evaluating electro-thermal response and degradation dynamics under realistic price variability. (c) Market prices conditional on dispatch and idle states on a semi-log plot, showing that the imposed dispatch rule preferentially samples higher-price regimes within the allowed window and therefore exposes assets to realistic revenue-relevant stress events without modeling explicit bidding or control logic.

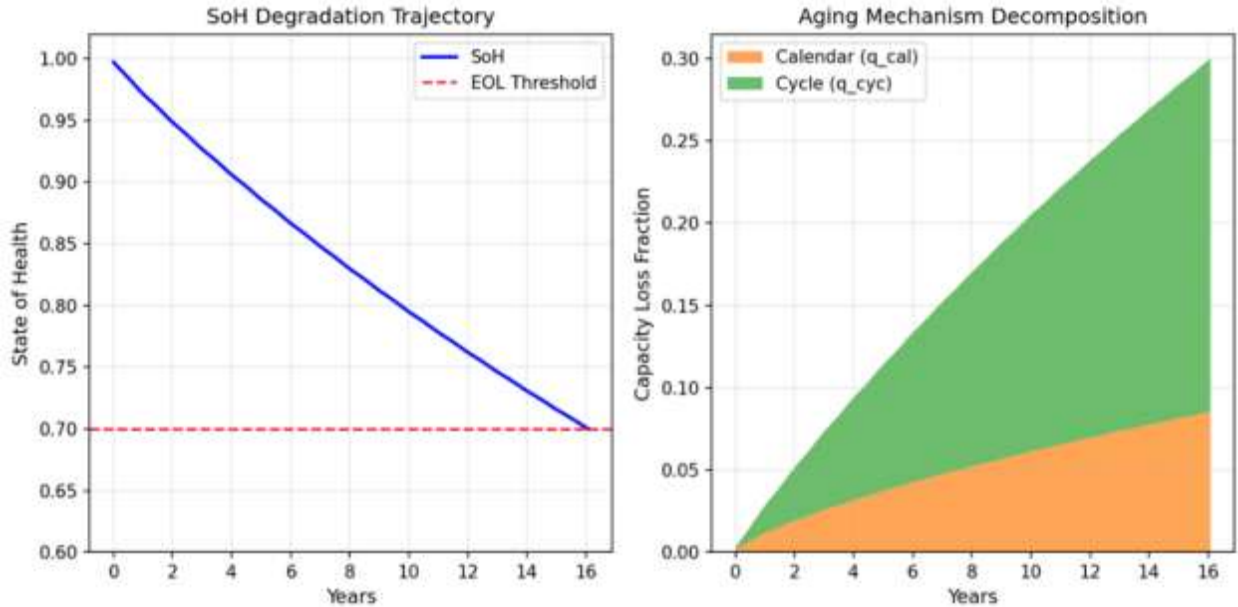

**Figure 3. Single-asset degradation trajectories and mechanism decomposition.** (a) State-of-health trajectory for a representative asset under daily arbitrage dispatch, reaching end-of-life (70% SOH) after approximately 20 years. The smooth concave-down profile reflects sublinear calendar aging combined with cumulative cycling effects. (b) Cumulative capacity loss decomposed into calendar aging and cycle aging contributions. Under the assumed operating profile, cycle aging accounts for the dominant fraction of total capacity loss by end of life, consistent with moderate cycling intensity.

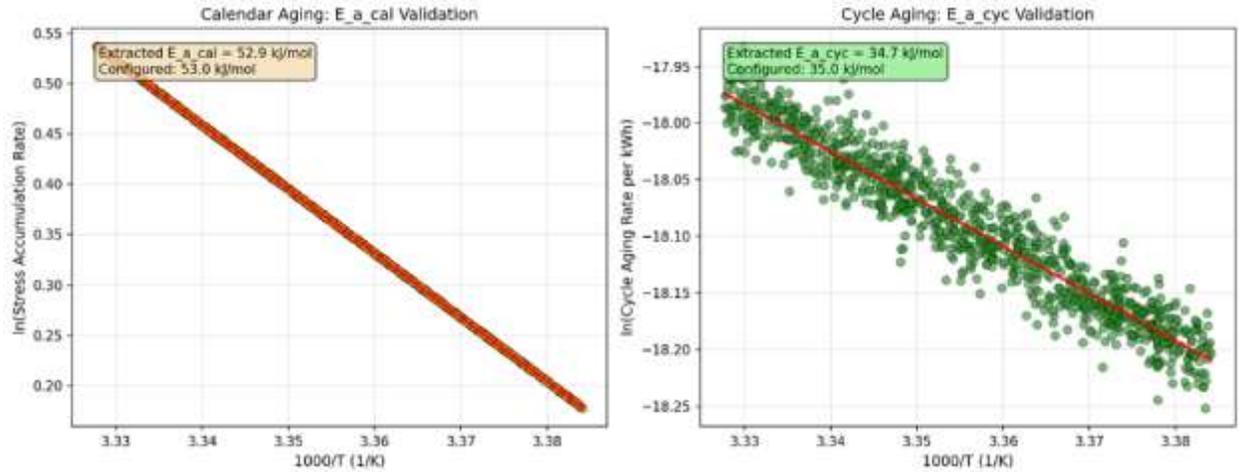

**Figure 4. Arrhenius consistency of thermal–degradation coupling.** (a) Calendar aging: natural logarithm of the stress accumulation rate versus inverse absolute temperature (1000/T). Linear regression yields an effective activation energy E_{a,cal}=52.9, in close agreement with the configured input value of 53. Minimal scatter reflects continuous accumulation of the Arrhenius-weighted stress metric and the absence of multiplicative quality-factor variability. (b) Cycle aging: natural logarithm of throughput-normalized capacity loss versus inverse absolute temperature. The extracted activation energy $E_{a,\text{cyc}}$ closely matches the configured input value. Increased scatter arises from intermittent accumulation during dispatch, lifetime temperature averaging, and inherited asset-to-asset quality variation

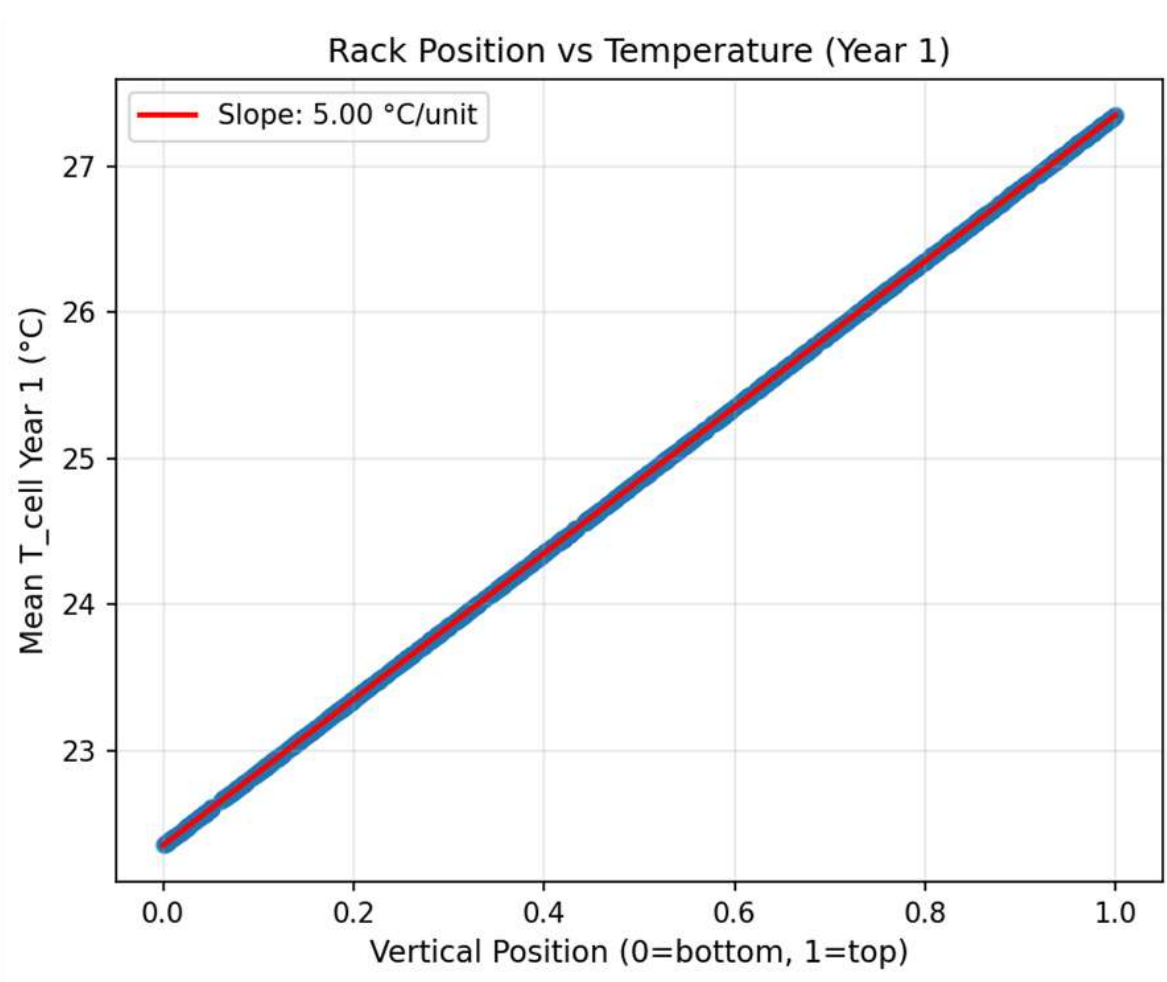

**Figure 5 | Thermal stratification drives fleet-level lifetime variability**. Mean cell temperature as a function of vertical rack position during the first year of operation, showing a persistent linear thermal gradient across the enclosure (slope ≈ 5 °C from bottom to top). This stratification is imposed by enclosure-scale thermal boundary conditions and remains stable over time.

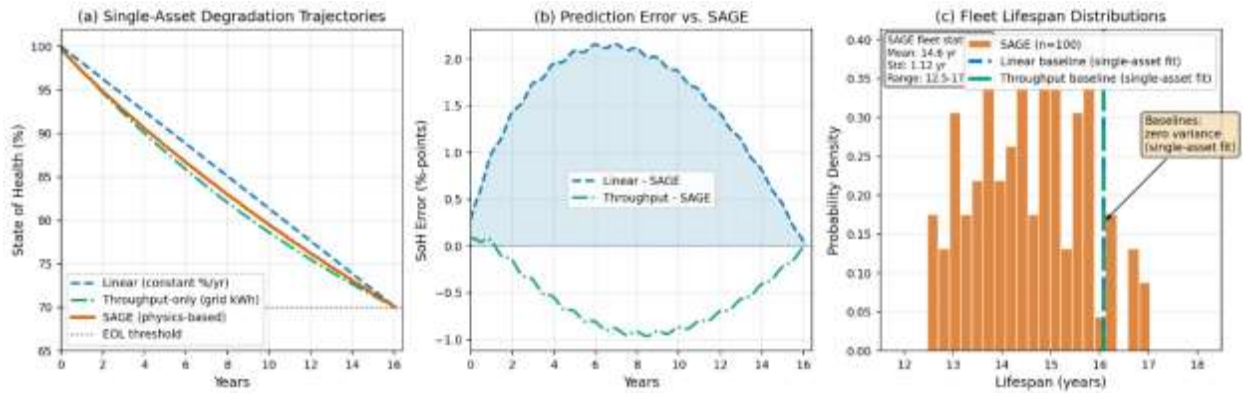

**Figure 6. Comparison of physics-based degradation with simplified baseline models.** (a) State-of-health trajectories for a representative asset predicted by a linear fade model, a throughput-only model, and the SAGE physics-based model for a single asset. All baseline models are calibrated to match the SAGE end-of-life time for the reference asset. The linear model predicts constant degradation, while the throughput-only model approximates the trajectory under the imposed regular dispatch conditions. (b) Prediction error relative to SAGE for the two baseline models, showing systematic mid-life error for the linear model and low error for the throughput-only model under steady cycling. (c) Fleet-level lifespan distributions obtained by applying each model to a heterogeneous asset ensemble. SAGE produces a finite distribution reflecting intrinsic heterogeneity in thermal environment and manufacturing quality, whereas the simplified baseline models collapse to a single deterministic lifespan.

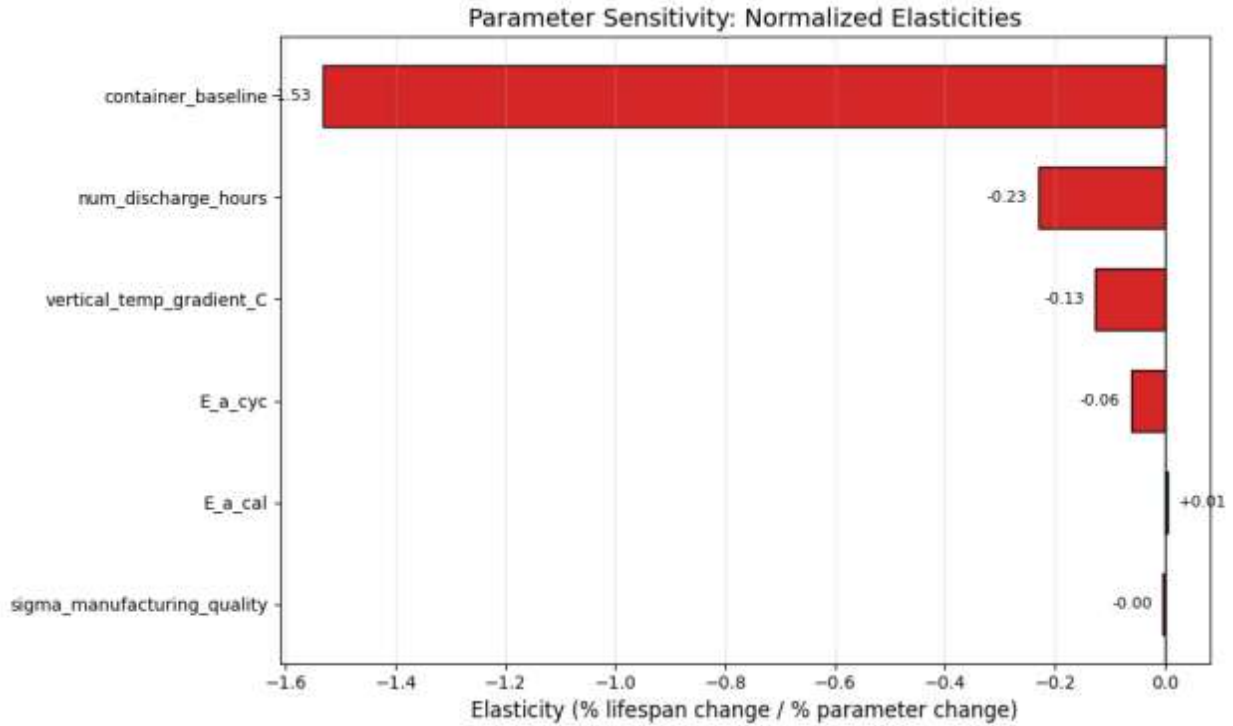

**Figure 7. Sensitivity of fleet-average lifespan to key model parameters.** Tornado diagram showing local elasticities of mean fleet lifespan with respect to selected SAGE model parameters, defined as the percent change in lifespan per percent change in parameter value about the baseline configuration. Parameters shown include container baseline temperature (HVAC-controlled setpoint), vertical thermal gradient within the enclosure $\Delta T_{\text{gradient}}$, cycle-aging activation energy $E_{a,\text{cyc}}$, calendar-aging activation energy $E_{a,\text{cal}}$, and manufacturing quality variability $\sigma_{\text{qual}}$. Negative elasticities (red) indicate that increasing the parameter reduces lifespan; positive elasticities (green) indicate lifespan extension. Results demonstrate that persistent container thermal conditions—particularly the HVAC-controlled baseline temperature and enclosure stratification—dominate fleet-average aging behavior, while kinetic degradation parameters and manufacturing variability exhibit comparatively weak influence on the mean under baseline thermal conditions.